\documentclass[referee]{aa} 

\def\ebv{\mbox{$\rm E(B-V)$}}
\def\ds{\mbox{$\rm d_\odot$}}
\def\rc{\mbox{$\rm R_{core}$}}

\begin{document}
\input{psfig}

%
\title{SOAR BVI Photometry of the Metal-poor Bulge Globular Cluster  NGC\,6642
\thanks{Observations collected with the SOAR Telescope, at Cerro Pachon, Chile } }
\author{
B. Barbuy\inst{1}
\and
E. Bica\inst{2}
\and
S. Ortolani\inst{3}
\and
C. Bonatto\inst{2}
}
%
%
\institute{
Universidade de S\~ao Paulo, Dept. de Astronomia, Rua do Mat\~ao 1226, 
S\~ao Paulo 05508-090, Brazil\\
 email: barbuy@astro.iag.usp.br
\and
Universidade Federal do Rio Grande do Sul, Dept. de Astronomia, 
CP 15051, Porto Alegre 91501-970, Brazil\\
 email: bica@if.ufrgs.br
\and
Universit\`a di Padova, Dipartimento di Astronomia, Vicolo
 dell'Osservatorio 2, I-35122 Padova, Italy\\
 email:  ortolani@pd.astro.it
}
\date{Received ; accepted }

\abstract{We present BVI  photometry of the globular
cluster NGC\,6642 using the SOI imager at the SOAR Telescope. The 
Colour Magnitude Diagrams (CMD) reach $\approx1.5\,$mag in V below the main sequence
turn-off. A comparison of the overall sequences, and in particular the
 Red Giant Branch slope of NGC\,6642 with that of M\,5
indicates that the two clusters must have a similar metallicity of
$\rm [Fe/H]\approx-1.3$.
We also obtain for NGC\,6642 a reddening E(B-V)=0.42$\pm$0.03, 
and a distance from the Sun
of d$_{\odot}$=7.2$\pm$0.5 kpc. 
Therefore NGC\,6642 is a moderately metal-poor globular 
cluster, spatially located in the bulge,  at a galactocentric
distance of $\rm R_{\rm GC}\approx1.7\,kpc$.
The comparison of CMDs of NGC\,6642 with those of M\,5 shows that  there
is a very good match of magnitude difference between turn-off and horizontal branch, 
suggesting comparable ages.  M\,5 has an age
typical of the halo globulars, and consequently NGC\,6642 is coeval with the halo.
 NGC\,6642 is a good candidate to be one of the few genuine
metal-poor and old {\it bulge} clusters, and might be one of the most
ancient fossils in the Galaxy. 
}
\titlerunning{SOAR Photometry of NGC\,6642}
\authorrunning{B. Barbuy et al.}
\maketitle
%

\section{Introduction}

For the vast majority of the known globular clusters in the Galaxy 
(Harris 1996,  as updated at
http://www.\-phy\-sics.mc\-mas\-ter.\-ca/Glo\-bu\-lar.\-html),
 the properties have already been inferred by means  of Colour-Magnitude Diagrams (CMD).
Until recently, no optical CMD was available for NGC\,6642.
In Barbuy et al. (1999) a revision of clusters within 20$^{\circ}$x20$^{\circ}$
around the Galactic center was presented, where NGC 6642 was not included 
due at the time to a lack of information on its HB morphology and 
distance to the Galactic center.

In Piotto et al. (2002),  74 CMDs in the HST WFPC2 F439W
and F555W bands were presented.  NGC\,6642 is included in that study, however
reddening and distance were adopted from Harris (1996). 
A JK CMD was presented by Minniti, Olszewski,
\& Rieke (1995), who  detected the cluster's Red Giant Branch
(RGB). Recio-Blanco et al. (2005) using Hubble data derived $\ebv=0.44$ and an
apparent distance modulus $\rm (m-M)_{F555W}=16.70$.

NGC\,6642, also designated ESO\,522-SC32 and GCL\,97,
is located at $\alpha_{2000}$ = 18$^{\rm h}$ 31$^{\rm m}$
54.3$^{\rm s}$, $\delta_{2000}$ = -23$^{\rm o}$ 28' 35"
($\ell = 9.81^{\circ}$, $b = -6.44^{\circ}$). 
It is  in Sagittarius, projected not far from the Galactic center.
Trager et al (1995) estimated 
a core radius  $\rc = 6.2\arcsec$, a concentration parameter $\rm c=1.99$,
which imply a tidal radius $\rm r_t\approx10\arcmin$.
The half-light radius is $\rm r_h= 44\arcsec$. 
Minniti (1995) derived $\rm [Fe/H] = -1.40$ from the spectroscopy
of 13 individual cluster stars.
The compilation by Harris (1996)
provides  $\rm [Fe/H] = -1.35$, $\ebv = 0.41$, $\ds = 7.7$ kpc.

In such central direction, bulge and inner halo are superimposed, and it is
important to derive accurate positions, kinematical data, metallicity and
abundance ratios
to characterize membership of globular clusters with respect to both Galactic subsystems.
Previous data indicate that NGC\,6642 is a metal-poor globular cluster located in the bulge,
and such borderline objects may provide clues on the bulge/inner halo issue, and in turn, on 
the early stages of the Galactic bulge formation. 

In this work we present deep BVI CMDs for NGC\,6642, deriving reddening, 
metallicity and distance. We also determine the age, for the first time for this cluster.

In Section 2 we describe the observations, data reduction and
calibration procedures.
In Section 3 we present the CMDs and measure 
cluster parameters. In Section 4 the relative age of 
NGC\,6642 is derived and discussed.
Concluding remarks are given in Section 5.

\section{Observations}

The SOAR  is a 4.1m telescope, located at Cerro Pachon, Chile,
and operated by AURA, for the consortium composed by 
CNPq-Brazil, NOAO, UNC and MSU. 

SOAR is presently being commissioned for the first time to scientific use.
The SOAR Optical Imager (SOI) is a bent-Cassegrain mounted optical 
imager using  two EEV $2050\times4100$ CCDs, 
 to cover a 5.26 arcminute square field of view at a scale of $\rm0.077\arcsec/pixel$. 

The observations of NGC\,6642 were carried out by the SOAR staff in 
June and July 2005 (Table~1) with 
the SOAR SOI camera in the B, V and I bands.
The full image has  a  gap of $10.8\arcsec$  between the 2 CCDs.
A binning $2\times2$ results in a pixel
size of $0.154\arcsec$.

\begin{table}
\caption[1]{\it Log of observations.}
\begin{flushleft}
\begin{tabular}{llllllll}
\noalign{\smallskip}
\hline
\noalign{\smallskip}
{\rm Target}& {\rm Filter} & {\rm Seeing (\arcsec)}&  {\rm Exp.Time [s]} & {\rm date} \\
\noalign{\smallskip}
\hline
\noalign{\smallskip}
${\rm NGC\;6642}$ & ${\rm B}$ & 0.9x0.8  & 15 & 06.06.2005 \\
 &      ${\rm B}$ &                    & 15 &  " \\  
 &     ${\rm B}$ &                     & 15 & "\\        
 &      ${\rm B}$ &                    & 15 & " \\
 &    ${\rm B}$ &            1.1x0.9 & 1080 & "\\
 &      ${\rm B}$ &            0.9x0.8 &  15  &    10.07.2005\\
 &      ${\rm B}$ &        "    &   "  &        "\\
 &      ${\rm B}$ &        "    &   "  & " \\
 &      ${\rm B}$ &        "    &   " & " \\
 &      ${\rm B}$ &        "    &   " & " \\
 &      ${\rm I}$ &      0.9x0.7 &  5 & " \\
 &      ${\rm I}$ &      0.78x0.75 & " & " \\
 &      ${\rm I}$ &      0.8x0.75 & " & " \\
 &      ${\rm I}$ &      0.8x0.7 & "& "  \\
 &      ${\rm I}$ &      0.8x0.7 & " & "  \\
 &      ${\rm V}$ &      0.8x0.7 & 7& "  \\
 &      ${\rm V}$ &      0.8x0.7 & "& "  \\
 &      ${\rm V}$ &      0.9x0.9 & "& "  \\
 &      ${\rm V}$ &      1.0x0.9&  " & "\\ 
 &      ${\rm V}$ &      0.8x0.75 & 600 & "\\ 
 &      ${\rm V}$ &      0.9x0.8 &  " & " \\
 &      ${\rm V}$ &      1.0x0.9 &  " & " \\
 &      ${\rm I}$ &      0.9x0.8 & 420  &   26.07.2005\\
\noalign{\smallskip} \hline \end{tabular}
\end{flushleft} 
\end{table}

The images were flatfielded, bias subtracted, trimmed, and
mosaiched by the SOAR Staff members. The photometry
was carried out using the DAOPHOT and ALLSTAR codes (Stetson 1994).

The absolute calibration was obtained from the standard stars in the region 
Markarian A (Landolt 1992)  observed on July 10th. 
This field is projected quite close to the 
cluster on the sky and it
contains standards spanning a wide range in colour (-0.24 $<$ $V-I$ $<$ 1.1).
The standard fields have been observed several times in BVI, 
collecting a total of 18 standard frames, almost all at the same airmass 
(1.19 - 1.21) in the night of July 10th. 
All the standard star images have been measured with MIDAS codes, 
using an aperture of 50 pixels, considerably larger than the seeing of 
FWHM (5-6 pixels), in order to avoid effects of 
frame to frame seeing variations. 
The very small instrumental magnitude variations of the same standards in different frames 
($<0.01$ mag.) indicate that the night was photometric.
The calibration equations, transforming the instrumental magnitudes (bvi) into calibrated 
magnitudes (BVI) obtained from these standards are:

$ V = v-0.08 (V-I)+28.60$

$ I = i+0.03(V-I)+27.77$

$ B = b+0.03(B-V)+29.19$

$ V = v-0.09(B-V)+28.60$

\noindent 
for 15s in $V$, 10s in $I$ and 25s in $B$ at 1.2 airmasses. The zero point errors
in the equations are about 0.01 mag. The first set of observations (June 6th) 
and the long exposure I image obtained on July 26th have been calibrated 
using the July 10th calibration, by transfer of 8 bright and isolated
stars in common.
Calibration of red stars is a well known problem because of lack of high quality,
very red  standards. Stars redder than $V-I\approx1.0 - 1.2$ are not only statistically 
rare, but often are long period variables, and only in a few cases they are heavily 
reddened normal stars (F-G type). When available, these stars often show a wide scatter 
about the linear extrapolation of the calibration from bluer stars. In this case
we prefer to use the extrapolation. In the present case, as usually occurs with
CCD photometric systems, the colour terms are rather small and, consequently the
calibrations do not depart much from the linear extrapolations,
 indicating that the passband system matches well the
 standard Johnson-Cousins system adopted in the Landolt standards.
We also checked the colour term calibration using our
previous observations obtained at the ESO Danish 1.5m telescope in 2000,
calibrated using Landolt (1992) standard stars from the T Phoenix field
containing standards with $V-I\leq1.65$. No significant colour deviation
was detected.

The photometric errors in NGC\,6642 photometry  can be obtained directly from 
DAOPHOT-Allstars outputs. The program gives the poissonian noise from the 
sky and star counts. The errors are less than 0.01 mag
between V=13 and 16, and reach 0.015 at V=17.5 in the short exposure V frames 
(7s). 

In the case of
crowded fields such as in NGC 6642, however, the poissonian noise is a lower 
limit indication of the real photometric error because it increases only as
a consequence of the higher background level. It does not take into account
the spatial noise (the residual noise from the flat fielding) neither the 
fitting errors induced by the blends. Extensive tests carried out in the past
demonstrated that the poissonian errors in globular cluster crowded fields are
typically 3-4 time lower than the errors derived from frame to frame or
from artificial star experiments.
For this reason we 
independently evaluated 
the photometric errors in the cluster area from frame to frame comparisons.
Using  $V$ images with exposure times of 7s and 600s,
errors of 0.015 mag have been derived at $V$=15, increasing up to 0.07 mag
at $V$=18, close to the limiting magnitude of the short exposure frame.
In principle these numbers should be representative of
errors in the short exposures because the signal to noise in the long exposure is 
considerably higher. A realistic evaluation of the  
photometric errors in crowded fields affected by differential reddening, 
is difficult to measure. Still the frame to frame method does not 
fully take into
account the blends because, if the seeing is not very different in the two frames, the blend 
effects are similar, but it takes into account the spatial noise and residual
defects of the detector (important mainly for bright stars) because the images are shifted 
by several pixels.
This means that the frame to frame errors, while more realistic of the poissonian error alone, 
still give a somewhat lower limit. 

In the subsequent analyses, we will employ CMDs derived from the combination of the short and 
long exposures, in order to avoid saturation effects.
 
Fig.~1 shows a 7 sec. $V$ image of NGC\,6642.

\begin{table}
\caption[1]{\it Variable stars in the direction of NGC\,6642.}
\begin{flushleft}
\begin{tabular}{llllllll}
\noalign{\smallskip}
\hline
\noalign{\smallskip}
{\rm Variable} & {$V$} & {$B-V$} & notes \\
\noalign{\smallskip}
\hline
\noalign{\smallskip}
 V1 &  13.1 &2.1 & long period \\
 V2 &  16.3 &1.2 &  type  non-identified \\
 V3 &  16.3& 0.6 &  RR Lyrae \\
 V4 &  16.6& 0.7 &  "  \\
 V5 &  16.7& 0.8&  "  \\
 V7 &  16.5 & 0.6& "  \\
 V9 &  16.4& 0.8& "  \\
 V11 & 16.2& 0.8& "  \\
 V12 & 16.2& 0.7& "  \\
 V13 & 16.6& 0.8& "  \\
 V15 & 16.3& 0.6 &   "  \\
\noalign{\smallskip} \hline \end{tabular}
\end{flushleft} 
\end{table}

\begin{figure}[ht]
\psfig{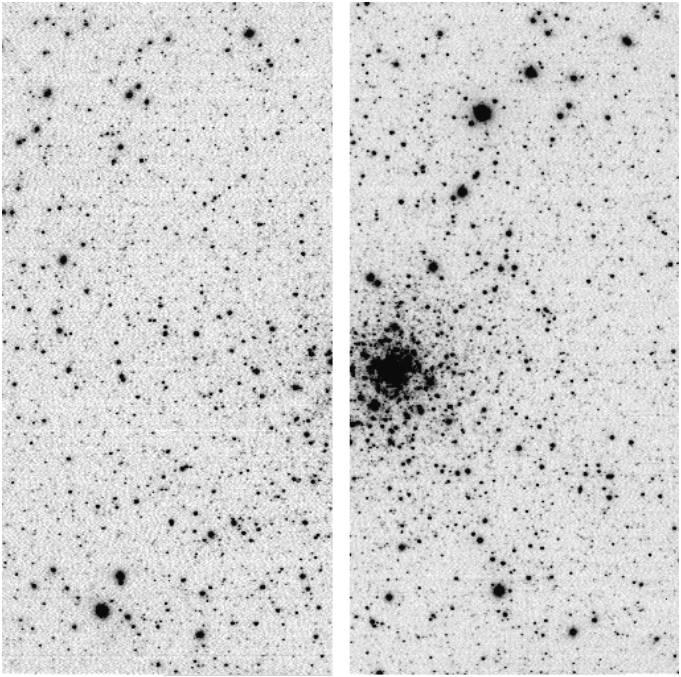}
\vspace{0cm}
\caption[ ]{
NGC\,6642 7 sec. $V$ image. Dimensions are 
5.26'$\times$5.26'. North is up and East is to the left.}
\label{n6642}
\end{figure}

\begin{figure}[ht]
\psfig{file=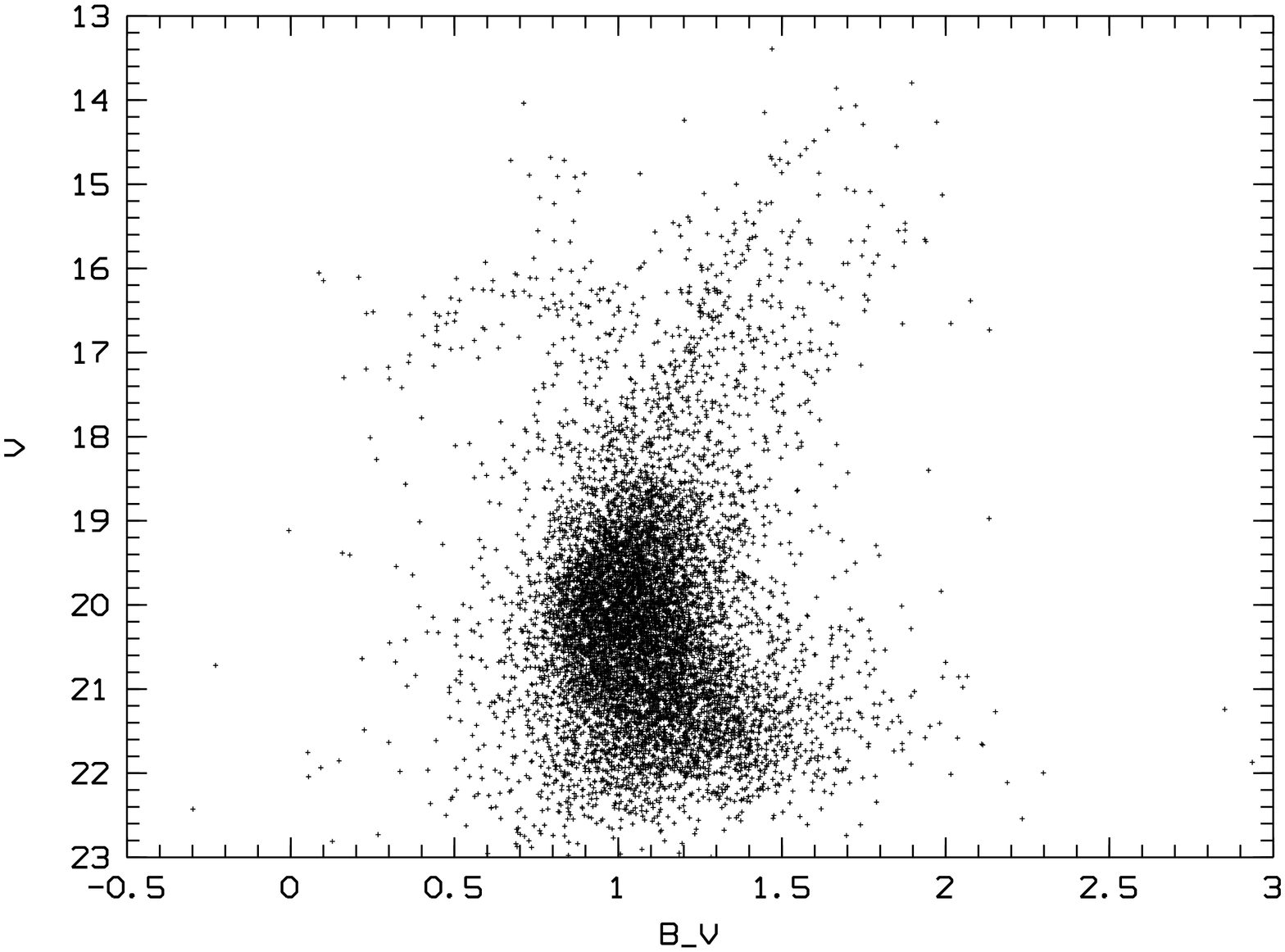,angle=-90,width=9.0 cm}
\psfig{file=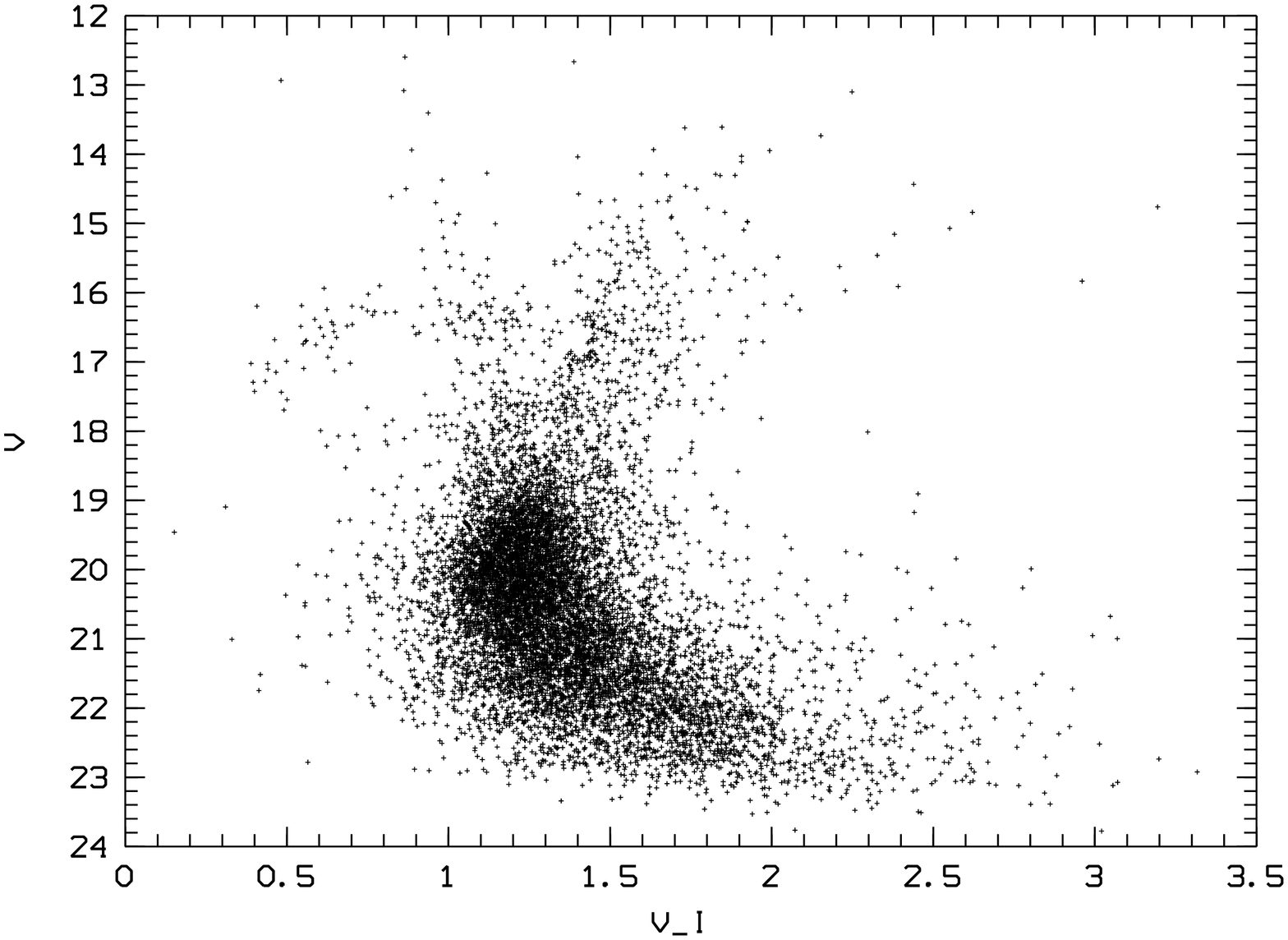,angle=-90,width=9.0 cm}
\vspace{0cm}
\caption[ ]{
$V$ vs. $B-V$ and $V$ vs. $V-I$ full field CMDs of NGC\,6642.}
\label{n6642full}
\end{figure}

\begin{figure}[ht]
\psfig{file=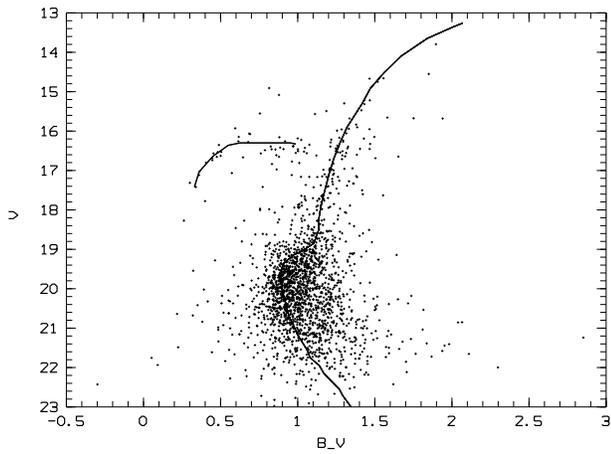,angle=-90,width=9.0 cm}
\vspace{0cm}
\caption[ ]{
$V$ vs. $B-V$ CMD of NGC\,6642, for an  extraction  
of $\rm0.13\leq r(\arcmin)\leq1.3$.
Mean locus of M\,5 CMD is overplotted,
 with shifts of $\Delta$V=1.28 and $\Delta$(B-V)=0.44.} 
\label{n6642vi}
\end{figure}

\begin{figure}[ht]
\psfig{file=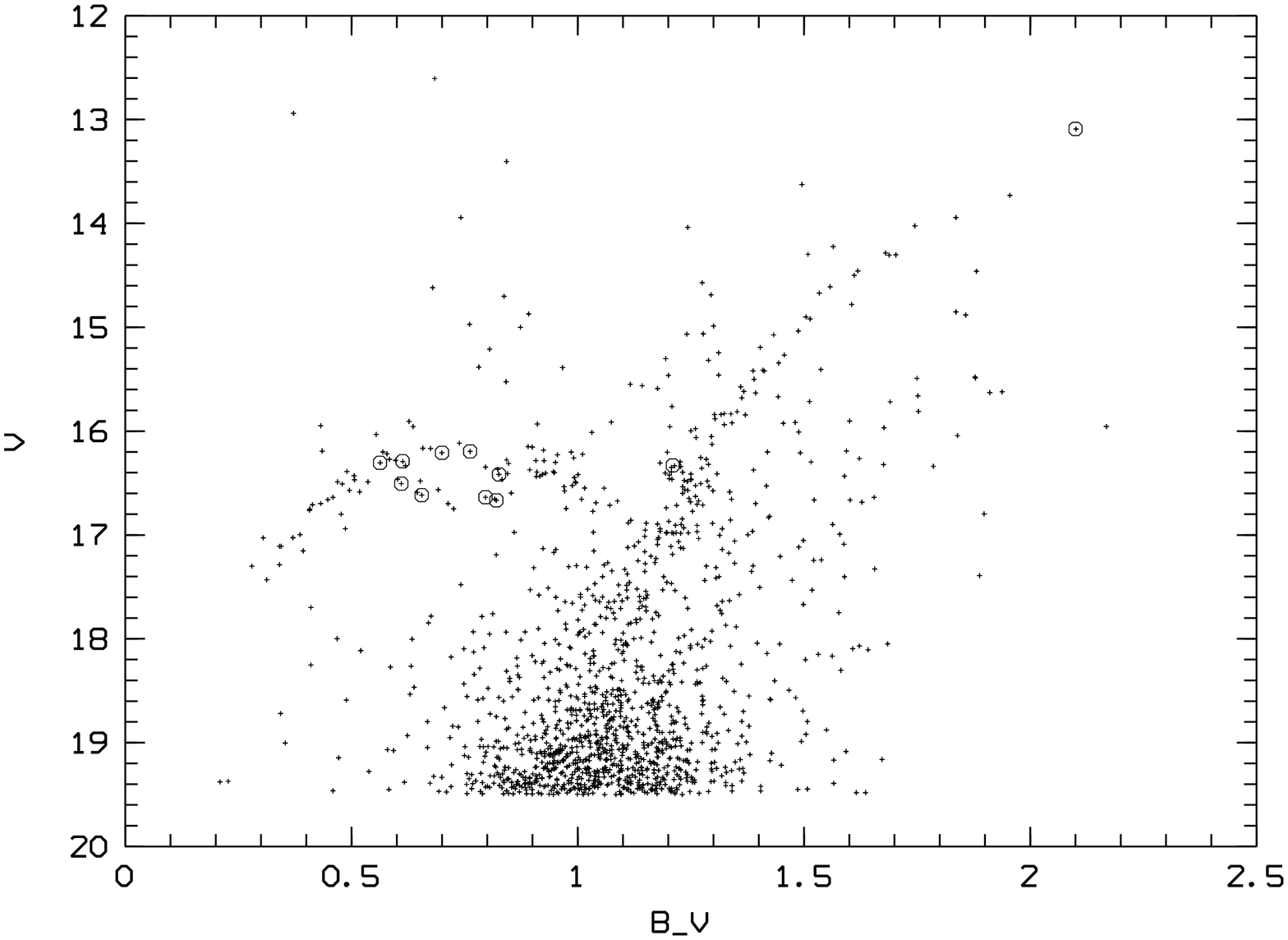,angle=-90,width=9.0 cm}
\vspace{0cm}
\caption[ ]{
$V$ vs. $B-V$ CMD of NGC\,6642 where variables are indicated.
Extraction is for r($\arcmin)\leq1.9$.}
\label{var}
\end{figure}

\section{Colour-Magnitude Diagrams}\label{cmd}

Figs. 2a,b show full field CMDs in $V$ vs. $B-V$ and $V$ vs. $V-I$.
 The horizontal branch (HB) is relatively rich in
stars and well defined.  The HB morphology includes blue and red stars
with respect to the RR-Lyrae gap.
Some Asymptotic Giant Branch (AGB) stars are
also present.

After trying to fit the mean loci of a number of template clusters of
different metallicities to the CMD of NGC\,6642, we found that the best
match is obtained using the CMD of M\,5 (NGC\,5904),  after applying
the appropriate shifts in magnitude and colour.
 In Fig.~3 we show the
$V$ vs. $B-V$ CMD for an extraction of 
$\rm 0.13\arcmin < R <1.3\arcmin$, where the mean locus of M\,5
from Johnson \& Bolte (1998) is overplotted.
The CMDs reach $\rm V\approx 21$ or almost $\approx1.5$\,mag  below the turn-off, 
which is located at $\rm V\sim19.7$.  

Fig.~3 shows that the upper evolutionary sequences (RGB, SGB and HB)
of NGC\,6642 and M\,5 are essentially coincident  (after applying
shifts in magnitude of $\Delta$V=1.28 and colour 
$\Delta$(B-V)=0.44, see below): 
in particular, the two RGBs
have the same slope.  Harris (1996) quotes a metallicity
$\rm [Fe/H] = -1.29$ for M\,5. The good match shown in Figure 3
indicates that NGC\,6642 must have $\rm [Fe/H]\approx -1.3$, confirming the
previous results of Minniti (1995).

The HB level is located at $\rm V_{HB}=16.35\pm0.04$ if we take the blue and red 
sides of the HB variable gap. The average value of the V magnitudes of the 9 
RR Lyrae we identified from Hazen (1993, hereafter H93) 
is $\rm V_{RR}=16.43\pm0.06$. It is not 
surprising that the average RR Lyrae magnitude is fainter because our
measurements are from very short exposure times as compared to the RR Lyrae
periods. Since the RR Lyrae have asymmetric light curves, with more time
spent at fainter magnitudes we expect their instantaneous luminosities to vary
from a minimum of 0.02 up to about 0.1 mag. fainter than the non variable HB stars.
The colour of the giant branch at the level of the HB is 
$B-V_{\rm RGB}$=1.26 and
$V-I_{\rm RGB}$=1.46.
For the reference cluster M\,5, the colours are  $B-V_{\rm RGB}$ = 0.85
 (Sandquist et al. 1996),
and   $V-I_{\rm RGB}$ = 0.97 (Johnson \& Bolte 1998). 
Therefore $\Delta$(B-V)=1.26-0.85=0.41 and $\Delta$(V-I)=1.46-0.97=0.51.
Given that the M\,5 reddening is E(B-V)=0.03 (Harris 1996) or E(V-I)=0.04,
we have a reddening of E(B-V)=0.44 and E(V-I)=0.54 for NGC\,6642.
The latter gives E(B-V)=0.54/1.33=0.41 using Dean, Warren \& Cousins (1978). 
The small difference in the reddening obtained from the two   
colours is likely due to differences in the zero points of the photometries.
Adopting an average of 0.42$\pm$0.03  and the total to selective absorption 
parameter 
R$_{\rm V}$=3.1 we get A$_{\rm V}=1.30\pm0.09$. The absolute distance modulus is 
(m-M)$_{\circ}$=16.35-1.30-0.74=14.3 (where the M$_{\rm V}$=0.74 has 
been adopted from Buonanno et al. 1989). Distance errors are dominated by 
uncertainties in the HB  level, $\rm\epsilon_V\approx0.1\,mag$. This error
includes the dispersion contributed by the instantaneous magnitudes of the RR Lyrae
stars. Considering in quadrature the errors of HB level and V absorption we
obtain a total error in distance modulus of $\pm0.13$. Accordingly, the 
distance from the Sun corresponds to d$_\odot$=7.2$\pm$0.5 kpc.

The Galactocentric coordinates of the cluster, assuming a distance of
the Sun to the Galactic center of $\rm R_{\odot} = 8.0$ kpc (Reid 1993),
 are X = -1.0
(X $<$ 0 is our side of the Galaxy), Y = 1.2, and Z = -0.8 kpc.  The
Galactocentric distance is $\rm R_{GC} \approx1.7$ kpc.  We conclude that
the cluster is spatially located within the bulge.

Concerning the kinematics, Harris (1996) gives a radial velocity $\rm v_{r,LSR}$= -45 km$\,s^{-1}$. 
This low velocity for NGC\,6642 in such central direction is compatible, within
uncertainties, with the bulge rotation (C\^ot\'e 1999). However, considering the
distribution of globular clusters in Fig.~11 of C\^ot\'e (1999), membership to the inner halo
cannot be ruled out. Another possibility is that NGC\,6642 is a halo cluster near perigalacticon, 
but in such  case a higher velocity would be expected. Proper motion determination
would nevertheless be necessary to rule out the halo alternative. In addition,
determination of metallicity and abundance ratios would provide further constraints on
the cluster membership to Galactic subsystems.

\subsection{Variables}
NGC\,6642 is known to contain many variable stars. H93 reported
the study of 18 RR Lyrae stars. Table 2 reports the magnitudes and colours
for 11 variables from H93 and they are indicated in Fig.~4.

We confirm that V1 is not a RR Lyrae star since it is located near the tip
of the giant branch. It has a too red colour and therefore it is
very likely a cluster long period variable, possibly Mira type for which 
Hoffleit (1972) gives P=216 days. The interpretation
of the red colour of V2 is more difficult. It is located on the RGB, 
with a considerably redder 
colour index than the RR Lyrae variable gap at 
0.60 $<$ $B-V$ $<$ 0.90. However both its
V magnitude (16.33) and its period measured by H93 (P=0.436 days) are
compatible with the cluster RR Lyrae. It is not easy to explain its
anomalous colour with blends or with a variable star of the field.
The remaining 9 variables are located in the expected cluster variable gap 
region from $B-V=0.61$ to 0.83 at an average magnitude of $V=16.43$, as already reported
in Section 3.
  As regards their location relative to the cluster center,
 V2 is quite close to the center of the cluster, located at about 30" from the center.
Half of the other variables we identified (V7, V13, V11, V12 and V3) are more distant.
 Therefore we confirm the previous H93's suggestion
that these variables belong to the cluster.
H93 obtained an average $B_{\rm RR} = 17.0$ and derived 
$V=16.3\pm0.2$ for the HB value, in very good agreement with our previous 
independent measurement of $V_{\rm HB}=16.35$. We cannot check directly the 
consistency of our
photometry with H93's zero point because we have no common secondary
standard, but there is no evidence for inconsistencies between the two
photometries.

\section{The Relative Age}


\begin{figure}[ht]
\psfig{file=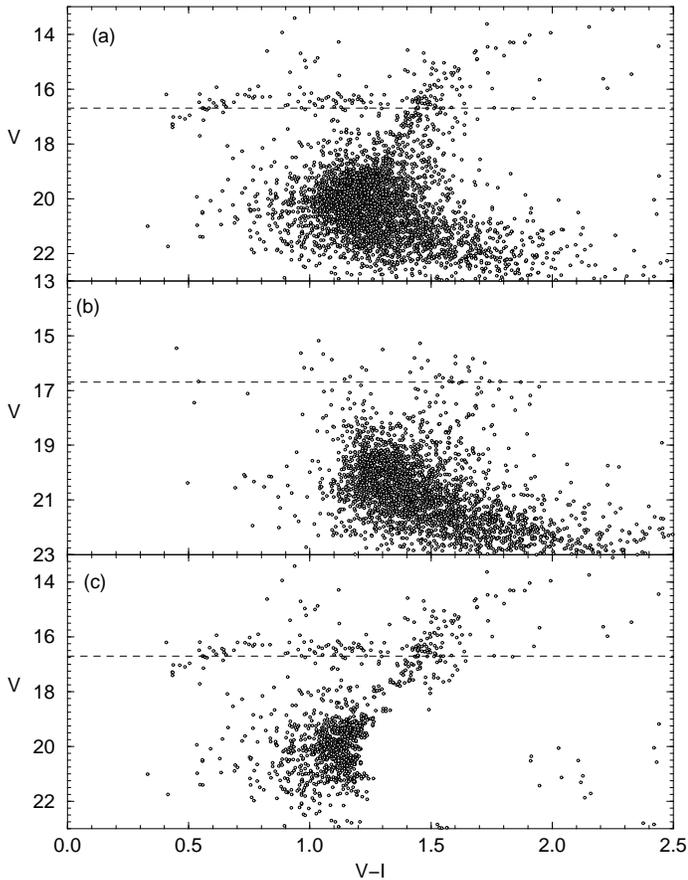,angle=0,width=9.0 cm}
\vspace{0cm}
\caption[ ]{
(a) - Original $V$ vs. $V-I$ CMD of the central region ($\rm 0.0-1.5\arcmin$) of NGC\,6642. 
(b) - Same area field-star CMD extracted at the East edge of the frame.
(c) - Decontaminated cluster CMD (see Sect. 4.1).
The dashed line shows the short/long exposure threshold.}
\label{clean}
\end{figure}

\begin{figure}[ht]
\psfig{file=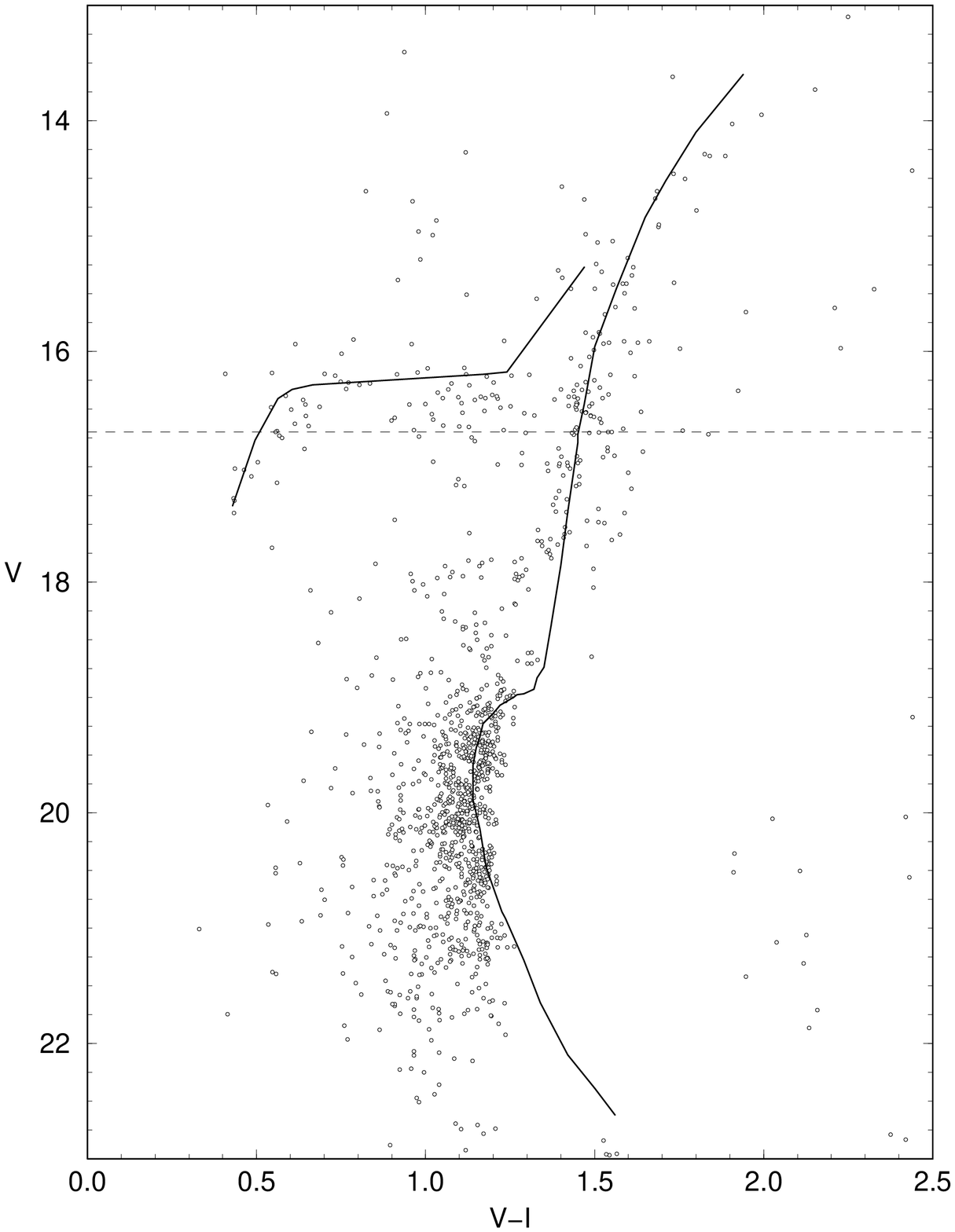,angle=0,width=9.0 cm}
\vspace{0cm}
\caption[ ]{
Same as Fig.5c where the mean locus of M\,5 CMD is overplotted.}
\label{n6642vi}
\end{figure}

\subsection{Field decontamination}

The relatively low latitude and bulge-intercepting line of sight of
NGC\,6642 imply that there is important field contamination, as shown by
the width of the Main Sequence (MS) and turn-off (TO) (Fig.~5a).
To minimize the effect of  foreground and
background field stars on the cluster CMDs we apply a decontamination
procedure based
on the number-density of stars present in the offset field.  As offset
field we use the North-South extension at the East edge of the CCD field, 
with size of 23 arcsec, this being the least cluster-contaminated frame zone.
This region contains a large enough number of
stars such as to produce a representative statistics of the field stars. Based
on the spatial number-density of stars in the offset field, the
decontamination algorithm estimates the number of field stars which
within the $\rm1\sigma$ Poisson fluctuation should be present
in the cluster field. The observed CMD is then divided into
colour/magnitude cells from
which stars are randomly subtracted in a number consistent with that
expected for
field stars in each cell. The dimensions of the colour/magnitude cells
can be subsequently changed so that the total number of stars subtracted
throughout the whole cluster area
matches the expected one, within the $\rm1\sigma$ Poisson fluctuation.
Since the field
stars are taken from an outer region of fixed dimensions, corrections
are made for the different solid angles of cluster and offset fields.
This procedure was previously used in the
analysis of low-contrast open clusters in the third quadrant (Bica \&
Bonatto \cite{BiBo2005}). 

The method is illustrated in Fig.~5 for an extraction of $\rm0.0\arcmin\leq R
\leq 1.5\arcmin$ in $V$ vs. $V-I$, where  the observed (panel (a)), same area
field-star (panel (b)), and decontaminated cluster CMDs (panel (c)) are shown. 
In all panels short and long exposures were combined. The dashed line shows the 
short/long exposure borderline. The giant branch and HB of NGC\,6642 are essentially not 
affected by field stars. On the other hand, in the original CMD (panel (a)) the subgiant 
branch of NGC\,6642 ($\rm V\approx18 - 19$) is considerably contaminated by field stars. 
In the decontaminated CMD (panel (c)) the subgiant branch is defined, although 
apparently somewhat depleted. This might be a real feature or a decontamination artifact.
A comparison with the HST CMD of the central parts of NGC\,6642 
(see Fig.~14 of Piotto et al. 
2002) shows similar CMDs, perhaps with the presence of a small gap at the
subgiant branch level. We note that the decontamination procedure was applied as
well to the bright range in Fig.5. However, the field-star density for bright stars
is so low compared to that of the cluster that essentially no star was subtracted. 

A fundamental result of the field subtraction (panel (c)) is the relatively narrow 
TO (within $V-I\approx1.0 - 1.2$), which in turn shows that bulge stars dominate the 
observed TO-red side (panel (a)). The decontaminated CMD helps constrain the cluster 
age (Sect.~4.2). 

\subsection{Age of NGC\,6642 relative to M\,5}

Fig.~3 shows the comparison of the dereddened $V$ vs. $B-V$ CMD of NGC\,6642 with the mean 
locus of the template cluster M\,5 (Sandquist et al. 1996). The upper sequences are very well 
reproduced, and by matching the blue and red parts of the HB and the GB, the M\,5 TO results 
coincident with the expected locus of NGC\,6642 TO.

Fig.~6 gives the $V$ vs. $V-I$ decontaminated CMD with the mean locus of M\,5
(Johnson \& Bolte 1998) overplotted,  with appropriate shifts in magnitude and colour. 
Similarly to the BV analysis, the bright sequences are well fitted and the
VI field subtracted TO is well reproduced in the range $V-I\approx1.0 - 1.3$, which
is considerably narrower than the distribution of stars in the BV CMD. The above 
indicates for NGC~6642 an age comparable to that of M~5. 
 We point out that at the 
RGB base some oversubtraction seems to have occurred. 
However the general fit is not
affected since the HB and GB extents provide constraints and leverage.

Rosenberg et al. (1999) and De Angeli et al. (2005) have shown that M\,5 has an age 
compatible with the mean age of the halo clusters. Therefore, NGC\,6642 is coeval with 
the halo.

\section{Concluding Remarks}

The SOAR telescope and the optical imager SOI have produced suitable scientific
images of NGC\,6642 in its first commissioning phase. Subarcsecond
images were obtained.

A CMD reaching below the turn--off was obtained, 
allowing to measure its age relative
to the template halo cluster M\,5. NGC\,6642 is coeval with M\,5, and therefore
with the halo.

The other parameters we obtained for NGC\,6642 are consistent
with the literature:  E(B-V) = 0.42$\pm$0.03, d$_{\odot}$= 7.2$\pm$0.5 kpc, 
and $\rm [Fe/H] = -1.3$.

It is interesting to note that this cluster shows an intermediate metallicity
in the tail of the metallicity distribution of the bulge (McWilliam \&
Rich 1994), and that it is spatially located within the bulge volume.

 It is important to note that the genuine bulge metal-poor globular clusters 
might be the most ancient fossil records of the Galaxy (van den Bergh 1993),
and  NGC\,6642 may be one of these objects.

Kinematical studies, in particular proper motions, would be of great
interest to verify to which component of the spheroid it belongs.
A further analysis of great interest would be to derive abundance ratios
from high resolution spectroscopy of individual stars, which might give
hints on characteristics of inner halo or bulge population for this cluster.

\begin{acknowledgements}
We acknowledge the grant Instituto do Mil\^enio - CNPq, 620053/2001-1.
 BB, EB and CB acknowledge partial financial 
support from the brazilian agencies CNPq and Fapesp.
 We thank Dr. Kepler de Oliveira for
helpful information and the SOAR staff for carrying out the observations
and pre-reducing the data.
The SOAR Telescope is operated by the Association
of Universities for Research in Astronomy, Inc.,
under a cooperative agreement between the CNPq, Brazil, 
the National Observatory
for Optical Astronomy (NOAO), 
University of North Carolina, and Michigan State University, USA.   
SO acknowledges the Italian Ministero dell'Universit\`a e della Ricerca
Scientifica e Tecnologica (MURST).

\end{acknowledgements}


%
%

%
\end{document}